\documentclass[ejs,preprint]{imsart}
\arxiv{stats.ME/1301.0463}
\usepackage{amssymb}
\usepackage{amsmath}
\usepackage{graphicx}
\usepackage{subfigure}
\usepackage{breakcites}
\usepackage[authoryear]{natbib}
\usepackage{epsfig,amsfonts,amsmath,enumerate,graphicx,amssymb,bm,color,amsthm}
\usepackage{hyperref}
\usepackage{algorithm}
\usepackage[]{algorithmic}
\usepackage{xspace}
\usepackage{mathtools,upgreek}
\usepackage{afterpage,lscape}
\usepackage[margin=1in]{geometry}

\newtheorem{theorem}{Theorem}

\newtheorem{corollary}{Corollary}

\newtheorem{remark}{Remark}

\begin{document}

\begin{frontmatter}
\title{Flexible linear mixed models with improper priors for longitudinal and survival data\thanksref{T1}}
\runtitle{Flexible linear mixed models}

\thankstext{T1}{This research was partially supported by EPSRC. We gratefully acknowledge insightful comment from two referees.}

\begin{aug}

\author{F. J. Rubio\ead[label=e1]{Francisco.Rubio@lshtm.ac.uk}}
\and
\author{M. F. J. Steel\ead[label=e2]{M.F.Steel@stats.warwick.ac.uk}}

\address{London School of Hygiene \& Tropical Medicine, London, UK.,\\\printead{e1}}
\address{University of Warwick, Department of Statistics, Coventry, CV4 7AL, UK.,\\\printead{e2}}

\runauthor{F. J. Rubio and M. F. J. Steel}
\end{aug}

\maketitle
\begin{abstract}
We propose a Bayesian approach using improper priors for hierarchical linear mixed models with flexible random effects and residual error distributions. The error distribution is modelled using scale mixtures of normals, which can capture tails heavier than those of the normal distribution. This generalisation is useful to produce models that are robust to the presence of outliers. The case of asymmetric residual errors is also studied. We present general results for the propriety of the posterior that also cover cases with censored observations, allowing for the use of these models in the contexts of popular longitudinal and survival analyses. We consider the use of copulas with flexible marginals for modelling the dependence between the random effects, but our results cover the use of any random effects distribution. Thus, our paper provides a formal justification for Bayesian inference in a very wide class of models (covering virtually all of the literature) under attractive prior structures that limit the amount of required user elicitation.
\end{abstract}



\tableofcontents
\end{frontmatter}

\section{Introduction}

Longitudinal and survival data appear in many application areas such as medicine, biology, agriculture, epidemiology, economics and small area estimation. This kind of observations typically present within- and among- subject dependence due to grouping and/or repeated measurements. Hierarchical linear mixed models (LMM) are often used to account for parameter variation across groups of observations as well as dependence. The general formulation of this type of model is given by
\begin{eqnarray}\label{LMM}
y_{ij} = {\bf x}_{ij}^{\top}\bm{\beta} + {\bf z}_{ij}^{\top} {\bf u}_i + \varepsilon_{ij},
\end{eqnarray}
where $y_{ij}$ is the observed response for subject $i$ at time $t_{ij}$, $j=1,\dots,n_i$ with $n_i$ recording the number of repeated measurements for subject $i$, and $i=1,\dots, r$ where $r$ denotes the number of subjects, $\bm{\beta}$ is a $p\times 1$ vector of \emph{fixed effects}, ${\bf u}_i$ are $q\times 1$ mutually independent random vectors (often called \emph{random effects}) and $\varepsilon_{ij}$ are \emph{i.i.d.} residual errors. Let ${\bf y}=\{y_{ij}\}$ be the $n\times 1$ vector of response variables and ${\bf X}$ and ${\bf Z}$ denote the known design matrices of dimension $n\times p$ and $n\times qr$, respectively, while $\bm{\varepsilon}$ is the $n\times 1$ vector of residual errors, and ${\bf u}=({\bf u}_1^{\top},\dots,{\bf u}_r^{\top})^{\top}$. We assume that $\operatorname{rank}({\bf X})=p$, and $n>p+qr$ throughout. In matrix notation we can write (\ref{LMM}) as
\begin{eqnarray*}
{\bf y} = {\bf X}\bm{\beta} + {\bf Z u} + \bm{\varepsilon}.
\end{eqnarray*}
These models are used in different fields under different names and notation. In Bayesian analysis of variance, the use of these models goes back to \cite{TT65}. In survival analysis, the $n$ survival times grouped in ${\bf T}$ are usually transformed to logarithms and modelled through
\begin{eqnarray}\label{MEAFT}
\log({\bf T}) = {\bf X}\bm{\beta} + {\bf Zu} + \bm{\varepsilon},
\end{eqnarray}
which is often referred to as a Mixed Effects Accelerated Failure Time model (MEAFT, \citealp{KL07}). In this literature, the random effects ${\bf u}$ are typically called \emph{frailties}. An additional challenge that often appears in this case is the presence of censored observations. In the context of production and cost frontiers used in economics, the logarithm of the output (or cost) of $r$ firms are modelled using (\ref{LMM}) with certain regularity restrictions on the regression coefficients $\bm{\beta}$ and sign restrictions on the random effects ${\bf u}_i$, which have a clear interpretation here in terms of inefficiencies. These models are known as stochastic frontier models (see \emph{e.g.}~\citealp{FOS97}).

The random vectors $\bm{\varepsilon}$ and ${\bf u}$ are often assumed to be normally distributed. Given that the normality assumption can be restrictive in practice (and even impossible for stochastic frontier models), alternative distributional assumptions have been explored, such as smooth flexible (semi-nonparametric) distributions for the random effects and normal residual errors \citep{ZD01}, skew Student-$t$ distributions for the random effects \citep{LT08}, normal random effects with residual errors modelled through a finite mixture of normals \citep{KL07,KL08}, and residual errors and random effects jointly distributed as a multivariate scale mixture of skew-normal distributions \citep{L10}, which makes the residual errors and the random effects dependent. \cite{J08} model both the residual errors and the random effects using multivariate skew--elliptical distributions, and \cite{D10} and \cite{J09} use Bayesian nonparametric approaches. Although flexible, Bayesian nonparametric approaches tend to be more computationally expensive than simpler flexible parametric models, while leading to similar conclusions as shown in our examples.

In a Bayesian framework with improper priors, \cite{HC96} and \cite{S01} present conditions for the propriety of the posterior under a certain improper prior structure for the parameters of model (\ref{LMM}) with normal assumptions on both the random effects and the residual errors. When using improper prior structures, it becomes essential to check the propriety of the posterior distribution in order to justify their use and interpretation in a probabilistic framework. \cite{HC96} characterise cases where certain improper prior structures used in practice lead to proper posteriors. The use of such priors was becoming common at that time since they do allow for the construction of a Gibbs sampler. However, \cite{HC96} show that the posterior may still be improper in some cases and, consequently, their use is not theoretically justified. \cite{R14} presents extensions of these propriety results to the use of certain classes of flexible random effects distributions. \cite{HC96} and \cite{R14} assume independence of the random effects, while \cite{S01} assume a specific structure for the covariance of the random effects. \cite{FOS97} proposed an improper prior structure which allows for using arbitrary random effects distributions, as long they are proper. More specifically, they adopt the following stochastic structure
\begin{eqnarray}\label{HierarchyFOS}
{\bf u} &\sim& p({\bf u}), \notag\\
\bm{\varepsilon} \mid \sigma_{\varepsilon} &\sim& \text{N}_n({\bf 0},\sigma_{\varepsilon}^2{\bf I}_n)\notag\\
\pi(\bm{\beta},\sigma_{\varepsilon}) &\sim& \dfrac{\pi(\bm{\beta})}{\sigma_{\varepsilon}^{b+1}}, \,\,\, b\geq 0,
\end{eqnarray}
where $p({\bf u})$ is proper, and $\pi(\bm{\beta})$ is proper or bounded. The condition that $p({\bf u})$ is proper is rather mild as this allows for using any parametric distribution for the random effects ${\bf u}$ with proper priors on the ``deeper'' parameters that index the random effects distribution. The propriety of $p({\bf u})$ is essential since an improper $p({\bf u})$ never leads to a posterior (see Theorem 2 from \citealp{FOS97}). The prior structure on $(\bm{\beta},\sigma_\varepsilon)$ is interesting as it includes priors with useful properties such as invariance with respect to reparameterisation under affine transformations.
In this paper, we  extend (\ref{HierarchyFOS}) by relaxing the assumption of normality of the residual errors and derive conditions for the existence of the corresponding posterior distribution that also allow for censored observations, covering virtually all models in the recent literature. Before presenting our results, we clarify our main contributions relative to our previous work in flexible linear regression models. \cite{RG16} study linear regression models with skew-symmetric errors and improper priors, covering the case with censored observations. Analogous results are presented in \cite{RY16} in the context of linear regression models with two-piece residual errors. Both papers focus on the effect of using skewed residual error distributions in terms of the prediction of the residual life of censored individuals. They do not study the propriety of the posterior distribution of linear regression models with random effects, which requires special attention as shown in this paper.

If not presented in the text, proofs are provided in the Supplementary Material (Appendix A).

\section{Multivariate scale mixtures of normals: a warning}\label{WarningCase}

In the context of linear regression, a common strategy for relaxing the assumption of normality of the residual errors consists in using multivariate scale mixtures of normals \citep{L10}. Recall that a $p$-variate scale mixture of normal distribution ($\text{SMN}_p(\bm{\mu},\bm{\Sigma},\delta;H)$) with location parameter $\bm{\mu}$, symmetric positive definite scale matrix $\bm{\Sigma}$, and parametric mixing distribution $H(\cdot\mid\delta)$ is defined through
\begin{eqnarray*}\label{MSMN}
f({\bf x} \mid \bm{\mu}, \bm{\Sigma},\delta) = \int_0^{\infty} \dfrac{\tau^{p/2}}{\det{(2\pi \bm{\Sigma})^{1/2}}} \exp\left[-\dfrac{\tau}{2}({\bf x}-\bm{\mu})^{\top}\bm{\Sigma}^{-1}({\bf x}-\bm{\mu})\right]dH(\tau\mid\delta).
\end{eqnarray*}
For example, when $H(\tau\mid\delta)$ is a Gamma distribution with parameters $(\delta/2,\delta/2)$ (and mean one), we obtain the $p$-variate Student-$t$ distribution with $\delta$ degrees of freedom. \cite{OS93} and \cite{B97} have pointed out inferential peculiarities using such error distributions, both in Bayesian and classical regression settings. Here we explore the implications of using multivariate scale mixtures of normal in the context of LMM.

Consider the following hierarchical structure for model (\ref{LMM})
\begin{eqnarray}
{\bf u}&\sim& p({\bf u}),\label{Hierarchy2}\\
\bm{\varepsilon}\mid \sigma_{\varepsilon},\delta_{\varepsilon} &\sim& \text{SMN}_n({\bf 0},\sigma_{\varepsilon}^2{\bf I}_n,\delta_{\varepsilon}; H_{\varepsilon}),\label{Hierarchy1}
\end{eqnarray}

\noindent where $p({\bf u})$ is proper. For our theoretical results, it does not matter whether $p({\bf u})$ is assumed to be a fixed distribution without further recourse to deeper parameters or whether it is a marginal distribution obtained from a proper joint distribution. In practice, the latter is typically the case and it is derived from a proper parametric distribution with a proper prior on its parameters. 
Throughout Sections \ref{WarningCase} and \ref{ProposedExtension}, we will only refer to $p({\bf u})$ for the sake of simplicity of notation.

We adopt the following prior structure
\begin{eqnarray}\label{Prior1}
\pi(\bm{\beta},\sigma_{\varepsilon},\delta_{\varepsilon}) &\sim& \dfrac{\pi(\bm{\beta}) \pi(\delta_{\varepsilon})}{\sigma_{\varepsilon}^{b+1}},
\end{eqnarray}
where $b\geq0$, $\pi(\bm{\beta})$ is bounded, and $\pi(\delta_{\varepsilon})$ is a proper prior with support $\Delta\subset {\mathbb R}$. The following result shows that the marginal likelihood of the data for this hierarchical model can be factorised as the product of the marginal likelihood under normality multiplied by a quantity that does not depend on the data.

\begin{remark}\label{FOSSMN}
For model (\ref{LMM}), (\ref{Hierarchy2})-(\ref{Prior1}) the marginal likelihood of the data can be written as
\begin{eqnarray}\label{FactorisationSMN}
\tilde{m}({\bf y}) &=& m({\bf y})\int_{\Delta}\int_{{\mathbb R}_+} {\tau}^{-\frac{b}{2}} \pi(\delta_{\varepsilon})d H_{\varepsilon}({\tau}\mid\delta_{\varepsilon}) d\delta_{\varepsilon},
\end{eqnarray}
where $m({\bf y})$ is the marginal likelihood corresponding to model (\ref{LMM}), (\ref{HierarchyFOS}).
\end{remark}

The factorisation of the marginal likelihood of the data can be used to obtain conditions for the propriety of the posterior distribution as follows.
\begin{corollary}\label{FOSSMNPP}
Consider the model (\ref{LMM}), (\ref{Hierarchy2})-(\ref{Prior1}) and let $(X:Z)$ denote the entire design matrix, and suppose that $p({\bf u})$ is proper. Consider also the following conditions:
\begin{enumerate}[(a)]
\item $\operatorname{rank}(X:Z)<n$,
\item $b\geq 0$,
\item  $\int_{\Delta}\int_{{\mathbb R}_+} \tau^{-\frac{b}{2}} \pi(\delta_{\varepsilon})d H_{\varepsilon}(\tau\mid\delta_{\varepsilon}) d\delta_{\varepsilon} < \infty$.
\end{enumerate}
Then, conditions (a) and (c) are necessary for the propriety of the posterior and conditions (a), (b), and (c) are sufficient for the propriety of the posterior.

\proof  From Remark \ref{FOSSMN} we have that the marginal density of the data can be written as in (\ref{FactorisationSMN}). The finiteness of $m({\bf y})$ in this expression, under assumptions (a)--(b), is proved in Theorem 1 from \cite{FOS97}. The second factor in (\ref{FactorisationSMN}) is finite by assumption (c).
\end{corollary}

For $b=0$, condition (c) is satisfied by any mixing distribution. However, for $b>0$ this condition may rule out the use of some mixing distributions and/or priors on $\delta_{\varepsilon}$, as discussed in the Supplementary Material (Appendix B). Remark \ref{FOSSMN} and Corollary \ref{FOSSMNPP} point out some issues with this  extension. If we use Bayes factors to compare two models ${\mathcal M}_1$ and ${\mathcal M}_2$ of the type described in Remark \ref{FOSSMNPP}, with different distributions for the residuals and corresponding marginal likelihoods $\tilde{m}_1({\bf y})$ and $\tilde{m}_2({\bf y})$, we obtain (using obvious notation)
\begin{eqnarray*}
B_{12} = \dfrac{\tilde{m}_1({\bf y})}{\tilde{m}_2({\bf y})} = \dfrac{\int_{\Delta}\int_{{\mathbb R}_+} {\tau}^{-\frac{b}{2}} \pi_1(\delta_{\varepsilon})d H_{\varepsilon, 1}({\tau}\mid\delta_{\varepsilon}) d\delta_{\varepsilon}}{\int_{\Delta}\int_{{\mathbb R}_+} {\tau}^{-\frac{b}{2}} \pi_{2}(\delta_{\varepsilon})d H_{\varepsilon, 2}({\tau}\mid\delta_{\varepsilon}) d\delta_{\varepsilon}}.
\end{eqnarray*}
Therefore, the Bayes factor is determined solely by the mixing distributions and their priors, but not by the data (!). Intuitively, these results indicate that the data do not contain information about the additional shape parameter $\delta_{\varepsilon}$. The factorisation (\ref{FactorisationSMN}) indicates that the use of this kind of multivariate extension is vacuous, in the sense that the additional shape parameters do not really help model selection. Similar issues with the use of certain classes of multivariate distributions and priors have been studied in \cite{MS14} for linear regression models. The next section presents an alternative extension that overcomes such problems.


\section{The proposed extension: flexible linear mixed models}\label{ProposedExtension}

The inferential issues pointed out in the previous section are due to the use of a single mixing variable in the construction of the joint distribution of the residual errors. In other words, \eqref{Hierarchy1} really implies only a single realisation of the mixing variable, irrespective of the size of $n$. Thus, we only have a single observation of the mixing variable, and inference on its (precision) parameter is impossible from the sample. One way to avoid these issues consists of using one mixing variable for each individual error. In this case, each observation contributes to the estimation of the parameter of the distribution of the mixing variable (see \citealp{W84} for a discussion in the context of Bayesian linear regression). Thus, we consider the  model in (\ref{LMM}) and (\ref{Hierarchy2}), but replace (\ref{Hierarchy1}) by
\begin{equation}\label{scalarSMN}
\varepsilon_{ij}\mid \sigma_{\varepsilon},\delta_{\varepsilon}, H_{\varepsilon} \stackrel{i.i.d.}{\sim} \text{SMN}_{1} ( 0, \sigma_{\varepsilon} ,\delta_{\varepsilon};H_{\varepsilon}).
\end{equation}
This is a rather wide class of symmetric and unimodal distributions, with tails which are either normal or fatter than normal, covering \textit{e.g.}~the normal, Student-$t$, logistic, Laplace, Cauchy and the exponential power family with power $1\le q<2$ (see \citealp{FS00} for more details). These distributions naturally lead to models that are robust to the presence of outliers (see \citealp{RPG03}, who also adopt $\text{SMN}_{1}$ residual errors, but restrict their study to the use of proper priors and normal random effects).

Using the same prior structure (\ref{Prior1}), we derive the following conditions for the propriety of the posterior distribution.

\begin{theorem}\label{FOSSMNPPExt1}
For the model (\ref{LMM}), (\ref{Hierarchy2}), (\ref{scalarSMN}) with prior (\ref{Prior1}), and $p({\bf u})$ a proper distribution, consider the following conditions:
\begin{enumerate}[(a)]
\item $\operatorname{rank}({\bf X}:{\bf Z})<n$,
\item $b\geq 0$,
\item  $\int_{\Delta}\int_{{\mathbb R}_+} \tau^{-\frac{b}{2}} \pi(\delta_{\varepsilon})d H_{\varepsilon}(\tau\mid \delta_{\varepsilon}) d\delta_{\varepsilon} < \infty$.
\item ${\bf y}$ is not an element of the column space of $({\bf X}:{\bf Z})$.
\end{enumerate}
Condition (a) is necessary for the propriety of the posterior. Conditions (a) -- (d) are sufficient for the propriety of the posterior.


\end{theorem}

Condition (a) indicates the need for repeated measurements (otherwise $n=r$ and $\operatorname{rank}({\bf X}:{\bf Z})=n$ for sensible choices of ${\bf Z}$), while conditions (b) and (c) characterise the priors and mixing distributions that produce a proper posterior. Given that all the distributional assumptions in Theorem \ref{FOSSMNPPExt1} are continuous, it follows that condition (d) is satisfied with probability one. Condition (d) is equivalent to saying that ${\bf y} = {\bf X}\bm{\beta} + {\bf Zu}$ has no solution, which can be checked numerically.

The only requirement on the distribution of the random effects ${\bf u}$ is that the marginal $p({\bf u})$ is proper, so as stated before, we can use an arbitrary parametric distribution ${\bf u}\sim F(\cdot\mid\bm{\theta})$, with a proper prior on the parameter $\bm{\theta}$. This includes, for example, the use of finite mixtures of normals, scale mixtures of normals, skewed scale mixtures of normals, truncated distributions, etc. In our examples, we explore the use of distributions that can capture departures from normality in terms of asymmetry and tail behaviour.

\subsection{Mixed effects accelerated failure time models with censoring}

The propriety result presented in Theorem \ref{FOSSMNPPExt1} can be extended to models used in survival analysis. Let ${\bf T}$ be a sample of $n$ survival times. The MEAFT in (\ref{MEAFT})
is exactly the same as (\ref{LMM}) for the log survival times. However, a common difficulty that arises with survival regression models is the presence of censored observations \citep{KL07,VS14}. 
If the sample of survival times does not contain censored observations, the existence of the posterior is provided by Theorem \ref{FOSSMNPPExt1}. If the sample contains both censored and uncensored observations, the propriety of the posterior can be based on the uncensored observations as follows.
\begin{theorem}\label{CenUncMEAFT}
Let ${\bf T}$ be a sample of survival times where $n_0$ observations are uncensored, with $p+qr<n_0<n$. Consider the model (\ref{MEAFT}), (\ref{Hierarchy2}), (\ref{scalarSMN}) with prior (\ref{Prior1}), and $p({\bf u})$ a proper distribution. Then, the posterior is proper if the marginal likelihood of the uncensored observations is finite.

\end{theorem}

The following result presents conditions for the case when the sample contains only censored observations.
\begin{theorem}\label{CenMEAFT}
Suppose that $n_c$ survival times $T_j$, $j=1,\dots n_c$, are observed as closed intervals $I_j$, and the rest of the observations exhibit another type of censoring. Thus, let $I_1,\dots,I_{n_c}$ be finite-length intervals on the positive real line, $n_c\leq n$ and let $({\bf X}:{\bf Z})_{n_c}$ represent the design matrix associated to the $n_c$ interval--censored observations. Consider the model (\ref{MEAFT}), (\ref{Hierarchy2}), (\ref{scalarSMN}) with prior (\ref{Prior1}), $p({\bf u})$ assumed to be proper, and the following condition:
\begin{enumerate}[(a)]
\item[(d')]  The set ${\mathcal E} = I_1 \times \dots \times I_{n_c}$ and the column space of $({\bf X}:{\bf Z})_{n_c}$ are disjoint. 
\end{enumerate}
Conditions (a)--(c) from Theorem \ref{FOSSMNPPExt1} together with (d') are sufficient for the propriety of the posterior.

\end{theorem}

Condition (d') means that there is no --unobserved-- true response that can be written as a linear combination of the columns of $X$ and $Z$, and can be relaxed in terms of the dimensionality as follows.
\begin{enumerate}[(a)]
\item[(\emph{d''})]  There exists a set ${\mathcal E}' = I_{j_1} \times \dots \times I_{j_{n'}}$, $p+qr<n' \leq n_c$, for some set of indexes ${\mathcal J}=\{j_1,\dots, j_{n'}\}\subseteq \{1,\dots,n\}$, such that ${\mathcal E}'$ and the column space of $({\bf X}',{\bf Z}')$ are disjoint, where ${\bf X}'$ and ${\bf Z}'$ are the design submatrices associated to the indexes ${\mathcal J}$.
\end{enumerate}
Another way of checking condition (d') consists of formulating it as a linear programming problem. Denote $\bm{\eta}\in {\mathbb R}^{p+qr}$, $\bm{\xi}=(\xi_1,\dots,\xi_{n_c})\in {\mathcal E}$, and $I_j = [l_j,u_j]$, $j=1,\dots,n_c$. Define the problem:
\begin{eqnarray}\label{LPP}
\text{Find}&& \max_{\bm{\eta},\bm{\xi}} 1,\notag\\
\text{Subject to}&& ({\bf X}:{\bf Z})_{n_c}\bm{\eta} = \bm{\xi},\notag\\
\text{and}&&\log(l_j)\leq \xi_j \leq \log(u_j),\,\,\, j=1,\dots,n_c.
\end{eqnarray}
Then, condition (d') is equivalent to verifying the infeasibility of (\ref{LPP}), for which several theoretical and numerical tools are available.

\subsection{Stochastic frontier models}

The stochastic frontier model with composed error takes the form of (\ref{LMM})
where ${\bf u}\in {\mathcal U}$ with ${\mathcal U}=\{{\bf u}: {\bf Z}{\bf u} \in {\mathbb R}_+^n\}$. Several specifications of interest in the context of economics are studied in \cite{FOS97}. The propriety results in Theorem \ref{FOSSMNPPExt1} apply to this model, which represents an extension of Theorem 1 in \cite{FOS97}.

\subsection{Asymmetric errors}

So far, we have considered symmetric extensions of the standard linear mixed model with normal errors. We now extend to asymmetric errors by using the class of two--piece distributions. Let $f$ be a continuous density with a unique mode at $0$, support on ${\mathbb R}$ and shape parameter $\delta$. A random variable $W$ is distributed according to a two-piece $f$-distribution (denoted $W\sim \text{TP}(\mu,\sigma,\gamma,\delta;f)$) if its density function can be written as:
\begin{eqnarray}\label{TPF}
s(w\mid \mu,\sigma,\gamma,\delta) = \dfrac{2}{\sigma[a(\gamma)+b(\gamma)]}\left[f\left(\dfrac{w-\mu}{\sigma b(\gamma)} \, \Big\vert \, \delta\right)I(w<\mu) + f\left(\dfrac{w-\mu}{\sigma a(\gamma)}\, \Big\vert \, \delta\right)I(w\geq\mu) \right].
\end{eqnarray}
where $I(\cdot)$ denotes the indicator function, and $\{a(\cdot),b(\cdot)\}$ are positive differentiable functions. This density is continuous, unimodal, with mode at $\mu\in{\mathbb R}$, scale parameter $\sigma\in{\mathbb R}_+$, and skewness parameter $\gamma\in\Gamma\subset\mathbb R$. It coincides with the density $f$ when $a(\gamma)=b(\gamma)$, and is asymmetric for $a(\gamma)\neq b(\gamma)$ while retaining the tails of $f$ in each direction. The most common choices for $a(\cdot)$ and $b(\cdot)$ correspond to the \emph{inverse scale factors} parameterisation $\{a(\gamma),b(\gamma)\}=\{\gamma,1/\gamma\}$, $\gamma \in {\mathbb R}_+$ \citep{FS98}, and the \emph{epsilon-skew} parameterisation $\{a(\gamma),b(\gamma)\}=\{1-\gamma,1+\gamma\}$, $\gamma \in (-1,1)$ \citep{MH00}. \cite{RS14a} study other parameterisations as well as some interpretable choices for the prior of the skewness parameter $\gamma$.

Suppose now that $f$ in (\ref{TPF}) is a univariate scale mixture of normals with mixing distribution $H_{\varepsilon}$ and consider the model (\ref{LMM}) with the following error structure
\begin{equation}\label{ISExtHierarchy1Asym}
\varepsilon_{ij}\mid \sigma_{\varepsilon},\gamma_{\varepsilon},\delta_{\varepsilon} \stackrel{i.i.d.}{\sim} \text{TP} ( 0, \sigma_{\varepsilon},\gamma_{\varepsilon} ,\delta_{\varepsilon};f),
\end{equation}
and prior
\begin{eqnarray}\label{PriorAsym}
\pi(\bm{\beta},\sigma_{\varepsilon},\gamma_{\varepsilon},\delta_{\varepsilon}) &\sim& \dfrac{\pi(\bm{\beta}) \pi(\gamma_{\varepsilon})\pi(\delta_{\varepsilon})}{\sigma_{\varepsilon}^{b+1}},
\end{eqnarray}
where $b\geq0$, $\pi(\bm{\beta})$ is bounded, and $\pi(\gamma_{\varepsilon})$ and $\pi(\delta_{\varepsilon})$ are proper priors with $\gamma_\varepsilon\in \Gamma.$

\begin{theorem}\label{FOSSMNPPExtAsym}
Consider the model (\ref{LMM}), (\ref{Hierarchy2}), (\ref{ISExtHierarchy1Asym}), (\ref{PriorAsym}), $p({\bf u})$ a proper distribution, and conditions (a)--(d) from Theorem \ref{FOSSMNPPExt1} together with the following condition:
\begin{enumerate}[(a)]
\item[(e)] $\int_{\Gamma} M(\gamma_{\varepsilon})^b\pi(\gamma_{\varepsilon}) d\gamma_{\varepsilon}<\infty$,\,\, where $M(\gamma_{\varepsilon})=\max\{a(\gamma_{\varepsilon}),b(\gamma_{\varepsilon})\}$.
\end{enumerate}
Then, conditions (a)--(e) are sufficient for the propriety of the posterior distribution.

\end{theorem}

Condition (e) is automatically satisfied for $b=0$ and any parameterisation $\{a(\cdot),b(\cdot)\}$. For the epsilon-skew parameterisation, condition (e) is satisfied for any $b\geq 0$ and any proper prior $\pi(\cdot)$ given that the functions $a(\cdot)$ and $b(\cdot)$ are upper bounded. Thus, the introduction of skewness does not destroy the existence of the posterior distribution while allowing for more flexibility.

\section{Numerical examples}\label{NumericalResults}

\subsection{Simulation study I (Student-$t$ errors, skewed random effects)}

In this section we conduct a simulation study, inspired by that presented in \cite{ZD01}, in order to assess the effect of different distributional assumptions on the random effects and the residual errors. We study the model:
\begin{eqnarray}\label{LMMSimulation}
y_{ij}= t_{ij}\beta_1 + w_i\beta_2 + u_i + \varepsilon_{ij},
\end{eqnarray}
\noindent where $i=1,\dots,r=100$, $j=1,\dots,n_i=5$, $t_{ij}=j-3$, $w_i=I(i\leq 50)$, $\beta_1=2$, $\beta_2=1$. According to \cite{ZD01}, $r=100$ represents a case where the amount of information available to estimate both the fixed effects and the random effects is modest. Four scenarios are considered for the distribution of the residual errors $\epsilon_{ij}$ and the random effects $u_i$. For the first scenario we simulate from $\epsilon_{ij}\sim \text{t}(0,0.5,2)$ and $u_i\sim \text{TPN}(-1.5,0.5,0.5)$; where $\text{t}(0,0.5,2)$ represents a Student-$t$ distribution with location parameter $0$, scale parameter $0.5$ and $2$ degrees of freedom, and $\text{TPN}(-1.5,0.5,0.5)$ represents a two--piece normal with location parameter $-1.5$, scale parameter $0.5$ and skewness parameter $0.5$ using the epsilon-skew parameterisation. For the second scenario we use $\epsilon_{ij}\sim \text{N}(0,0.5)$ and $u_i\sim \text{TPN}(-1.5,0.5,0.5)$. The third scenario consists of $\epsilon_{ij}\sim \text{t}(0,0.5,2)$ and $u_i\sim \text{N}(-1.5,0.5)$. The fourth scenario uses $\epsilon_{ij}\sim \text{N}(0,0.5)$ and $u_i\sim \text{N}(-1.5,0.5)$. We simulate $100$ data sets under these configurations. For each of these simulated samples, the model (\ref{LMMSimulation}) was fitted assuming that $\epsilon_{ij}\sim t(0,\sigma_{\varepsilon},\delta_{\varepsilon})$ and $u_i \sim \text{TPN}(\mu,\sigma,\gamma)$ with the prior structure:
\begin{eqnarray}\label{PriorSimulation}
\pi(\beta_1,\beta_2,\sigma_{\varepsilon},\delta_{\varepsilon},\mu,\sigma,\gamma) \propto \dfrac{\pi(\delta_{\varepsilon})\pi(\mu)\pi(\sigma)\pi(\gamma)}{\sigma_{\varepsilon}},
\end{eqnarray}
\noindent where $\pi(\delta_{\varepsilon})$ is the weakly informative priors for the degrees of freedom of a Student-$t$ distribution proposed in \cite{RS14b}, $\pi(\mu)$ is a uniform prior on $[-100,100]$, $\pi(\sigma)$ is a half-Cauchy density with mode $0$ and unit scale parameter, which has been shown to induce a posterior with good frequentist properties in the context of Bayesian hierarchical models \citep{PS12}. Finally, $\pi(\gamma)$ is a uniform prior on $(-1,1)$, which represents a weakly informative prior as described in \cite{RS14a}. The prior in (\ref{PriorSimulation}) is consistent with prior (\ref{Prior1}) and ensures a proper $p(u_i)$. The propriety of the posterior is then guaranteed by Theorem \ref{FOSSMNPPExt1}. For each of the $100$ simulated sets we obtain a posterior sample of size $1,000$ using an adaptive Metropolis within Gibbs algorithm \citep{RR09} using the `spBayes' R package \citep{F07}. The posterior samples are obtained after a burn-in period of $7500$ iterations and thinned every $10$ iterations in order to reduce correlation ($17,500$ draws in total). Table \ref{table:SimulationGM} presents summary statistics of the posterior samples, the median (over the datasets) Bayes factors in favour of $\gamma=0$ (symmetric random effects), calculated using the Savage-Dickey density ratio, and the average odds $p/(1-p)$, where $p={\mathbb P}(\delta_{\varepsilon}>10)$, which provides information about the appropriateness of the assumption of normal tails for the residual errors. The Savage-Dickey density ratio \citep{VW95} is calculated as the ratio of the marginal posterior density function and the prior density of the parameter under consideration ($\gamma$) evaluated at the point of interest ($\gamma=0$). In order to evaluate the posterior density, we employ a kernel density estimator (with Gaussian kernel) of the posterior samples. The use of the (standard) Savage-Dickey density ratio is valid in our setting as we are using independent proper priors for the parameters that differ across models (see \citealp{VW95}). Posterior medians are presented instead of posterior means as the existence of the posterior mean is not guaranteed in all the scenarios. From this table, we can observe that despite the relatively small sample size, these Bayesian model selection criteria correctly identify the true model in each scenario.

In order to assess the impact of the assumption of normality, we implement the model with normal residual errors and random effects. Table \ref{table:SimulationNM} summarizes posterior inference for this model. The misspecification of the distribution of the residual errors leads to an overestimation of $\sigma_{\varepsilon}$, which is in line with the fact that this scale parameter has a different interpretation in normal and Student-$t$ cases, while the presence of unmodelled skewness affects the estimation of the location parameter $\mu$. Note also that misspecification dramatically increases the uncertainty about the fixed effects.

\begin{table}[h!]
\small
\begin{center}
\begin{tabular}[h]{|c|c|c|c|c|}
\hline
 & Scenario 1  & Scenario 2 & Scenario 3 & Scenario 4  \\
Parameter &  $\varepsilon_{ij}\sim t$ & $\varepsilon_{ij}\sim \mbox{N}$ & $\varepsilon_{ij}\sim t$  & $\varepsilon_{ij}\sim \mbox{N}$  \\
 & $u_i\sim \mbox{TPN}$  & $u_i\sim \mbox{TPN}$  & $u_i\sim \mbox{N}$  & $u_i\sim \mbox{N}$  \\
\hline
$\beta_1$ &  1.99 (1.96,2.04) & 2.00 (1.98,2.03)  & 2.00 (1.96, 2.03)  &  2.00  (1.98,2.03)   \\
\hline
$\beta_2$  & 0.99 (0.74,1.23) & 0.99 (0.80,1.20)  &  1.01 (0.75, 1.21)  &  1.00 (0.76,1.19)   \\
\hline
$\sigma_{\varepsilon}$  & 0.50 (0.42,0.57) & 0.48 (0.44,0.52)  & 0.50 (0.42,0.57)  &  0.48 (0.44,0.52)   \\
\hline
$\delta_{\varepsilon}$  & 1.97 (1.61,2.71) &  31.54 (13.64, 203.5)  & 1.96 (1.60,2.70)  &   30.48 (13.73,203.4)   \\
\hline
$\mu$   & -1.50 (-1.97, -1.20) & -1.50 (-1.92,-1.22)  & -1.47 (-1.91,-1.04)  &   -1.50 (-1.95,-1.11)   \\
\hline
$\sigma$  & 0.49 (0.38,0.60) &  0.49(0.41,0.57)  & 0.48 (0.37,0.59)  &  0.49 (0.41,0.59)   \\
\hline
$\gamma$  & 0.50 (0.00,0.79) & 0.54 (-0.04,0.79)  & 0.00 (-0.51,0.58)  &   0.01 (-0.52,0.41)   \\
\hline
Median BF  & & & & \\ $\gamma=0$ &  0.78   &  0.43  &  1.98  &  2.50  \\
\hline
Median odds & & & & \\ $\delta_{\varepsilon}>10$  &  0 &  17.54  & 0 &   19.02  \\
\hline
\end{tabular}
\caption{ Simulation study I: Posterior medians and 95\% credible intervals of the median estimators using the general model. The median odds are calculated after removing the cases where all of the drawn values are above $10$.}\label{table:SimulationGM}
\end{center}
\end{table}

\begin{table}[h!]
\small
\begin{center}
\begin{tabular}[h]{|c|c|c|c|c|}
\hline
 & Scenario 1  & Scenario 2 & Scenario 3 & Scenario 4  \\
Parameter &  $\varepsilon_{ij}\sim t$ & $\varepsilon_{ij}\sim \mbox{N}$ & $\varepsilon_{ij}\sim t$  & $\varepsilon_{ij}\sim \mbox{N}$  \\
 & $u_i\sim \mbox{TPN}$  & $u_i\sim \mbox{TPN}$  & $u_i\sim \mbox{N}$  & $u_i\sim \mbox{N}$  \\
\hline
$\beta_1$ & 2.00 (1.89,2.09) & 2.00 (1.98, 2.03)  &  2.00 (1.89, 2.10)  & 2.00 (1.97,2.02)     \\
\hline
$\beta_2$ & 0.99 (-0.05,2.24) & 1.07 (0.13,2.39)  &  1.08  (-0.05,2.05)  & 1.01 (0.78,1.20)    \\
\hline
$\sigma_{\varepsilon}$ &  1.53 (1.01,3.05) & 0.50 (0.47,0.53)  &  1.53 (1.01,3.05)  & 0.50 (0.46,0.53)    \\
\hline
$\mu$  & -3.41 (-4.21,-2.82) & -3.51 (-4.28,-2.97)  &  -1.53 (-2.18,-0.92)  & -1.50 (-1.65,-1.35)  \\
\hline
$\sigma$ &  2.57 (2.18,3.00) & 5.27 (2.24,2.94)  &  2.44 (2.07,2.86)  & 0.50 (0.42,0.59)     \\
\hline
\end{tabular}
\caption{ Simulation study I:  Monte Carlo medians and 95\% credible intervals of the median estimators using the normal model}\label{table:SimulationNM}
\end{center}
\end{table}


\subsection{Simulation study II (Student-$t$ errors and random effects with censoring)}\label{Sec:SimII}

In this section, we present a simulation study with censored responses. We consider a model which is used in practice for modelling the evolution of a marker in longitudinal studies (see \emph{e.g.}~\citealp{VL09}), in particular
\begin{eqnarray*}
y_{ij} = \beta_j + u_i + \varepsilon_{ij},
\end{eqnarray*}
where $\varepsilon_{ij}$ and $u_i$ are mutually independent, $i=1,\dots,100$, and $j=1,\dots,5$. We generate data using two scenarios: (I) $\varepsilon_{ij}\stackrel{i.i.d.}{\sim} N(0,0.5)$, and $u_i \stackrel{i.i.d.}{\sim} N(0,0.25)$; and (II) $\varepsilon_{ij}\stackrel{i.i.d.}{\sim} t(0,0.5,2)$, and $u_i \stackrel{i.i.d.}{\sim} t(0,0.25,2)$. The theoretical values of the regression parameters are $(\beta_1,\beta_2,\beta_3,\beta_4,\beta_5)=(2.5,3.0,3.5,4.0,4.5)$.  We simulate 100 data sets under these two configurations and truncate the values that are greater than $\log_{10}(75000)$. This truncation strategy produces samples with 7\%--18\% censored observations. Then, we fit the following four models:  Model 1, $\varepsilon_{ij}\stackrel{i.i.d.}{\sim} N(0,\sigma_{\varepsilon})$, and $u_i\stackrel{i.i.d.}{\sim}N(0,\sigma)$, Model 2, $\varepsilon_{ij}\stackrel{i.i.d.}{\sim} t(0,\sigma_{\varepsilon},\delta_{\varepsilon})$, and $u_i\stackrel{i.i.d.}{\sim}N(0,\sigma)$, Model 3, $\varepsilon_{ij}\stackrel{i.i.d.}{\sim} N(0,\sigma_{\varepsilon})$, and $u_i\stackrel{i.i.d.}{\sim}t(0,\sigma,\delta)$ and Model 4, $\varepsilon_{ij}\stackrel{i.i.d.}{\sim} t(0,\sigma_{\varepsilon},\delta_{\varepsilon})$, $u_i\stackrel{i.i.d.}{\sim} t(0,\sigma,\delta)$. For the most general model (Model 4) we adopt the prior structure:
\begin{eqnarray}\label{PriorSimulationII}
\pi(\bm{\beta},\sigma_{\varepsilon},\delta_{\varepsilon},\sigma,\delta) \propto \dfrac{\pi(\delta_{\varepsilon})\pi(\sigma)\pi(\delta)}{\sigma_{\varepsilon}},
\end{eqnarray}
\noindent where $\pi(\delta_{\varepsilon})$ and $\pi(\delta)$ are the weakly informative priors for the degrees of freedom of a Student-$t$ distribution proposed in \cite{RS14b}, $\pi(\sigma)$ is a half-Cauchy density with mode $0$ and unit scale parameter. For Models 1-3 we use the obvious reduction of this prior. The propriety of the posterior is then guaranteed by Theorem \ref{CenUncMEAFT}.

Tables \ref{table:SimulationI} and \ref{table:SimulationII} (with a star indicating the model used to generate the data) give a summary of the posterior results.

\begin{table}[h!]
\small
\begin{center}
\begin{tabular}[h]{|c|c|c|c|c|}
\hline
Parameter & Model 1*  & Model 2 & Model 3 & Model 4  \\
\hline
$\beta_1$ &   2.49 (2.37,2.61) &   2.50  (2.37,2.61) &  2.50  (2.38,2.61) &   2.49  (2.37,2.61)    \\
\hline
$\beta_2$  &  2.99 (2.90,3.08) &  2.99  (2.91,3.08) &    2.99 (2.91,3.07) &    2.98  (2.90,3.07)    \\
\hline
$\beta_3$  &  3.51 (3.39,3.60) &  3.51  (3.39,3.60) &   3.51  (3.39,3.60) &  3.51 (3.39,3.60)    \\
\hline
$\beta_4$  &  3.99 (3.90,4.14) &  3.99  (3.90,4.14) &  3.99  (3.89,4.14) &  4.00 (3.90,4.14)    \\
\hline
$\beta_5$  & 4.49  (4.40,4.62) &   4.50  (4.41,4.63) &  4.49  (4.40,4.63) &  4.49   (4.40,4.62)    \\
\hline
$\sigma_{\varepsilon}$  & 0.50  (0.47,0.54) &   0.49  (0.46,0.53) &  0.50 (0.47,0.54) &   0.48 (0.44,0.53)    \\
\hline
$\delta_{\varepsilon}$  &   -- &   59.51  (29.15,201.12) &    -- &  32.73  (11.28,212.87)    \\
\hline
$\sigma$  &  0.25 (0.20,0.31) &  0.25  (0.20,0.31) &  0.22  (0.15,0.28) &   0.22  (0.16,0.28)    \\
\hline
$\delta$  &   -- &    -- &   11.33 (4.64,97.28) &   10.77  (4.41, 18.78)    \\
\hline
\end{tabular}
\caption{ Simulation study II with Scenario I: Posterior medians and 95\% credible intervals of the median estimators.}\label{table:SimulationI}
\end{center}
\end{table}

Table \ref{table:SimulationI} suggests that using a more complex model than necessary does not affect the inference on the parameters in the simple model (Model 1), which is encouraging, as we do not really seem to lose anything by allowing for flexibility. We merely find out that the assumed Student-$t$ distributions have large degrees of freedom parameters, which makes them close to normal.

\begin{table}[h!]
\small
\begin{center}
\begin{tabular}[h]{|c|c|c|c|c|}
\hline
Parameter & Model 1 & Model 2 & Model 3 & Model 4*  \\
\hline
$\beta_1$ &   2.45  (2.06,2.67) &  2.49 (2.33,2.68) & 2.49 (2.18,2.69) &  2.50 (2.35,2.66)    \\
\hline
$\beta_2$  &  2.96  (2.55,3.26) &  3.00 (2.83,3.22) & 3.00 (2.65,3.34) &  3.00 (2.85,3.20)    \\
\hline
$\beta_3$  &   3.43  (3.08,3.70) &  3.48 (3.31,3.68) & 3.47 (3.22,3.70) &   3.49 (3.34,3.65)    \\
\hline
$\beta_4$  & 3.98  (3.68,4.21) &  4.00 (3.80,4.20) & 4.00 (3.71,4.24) & 4.01  (3.82,4.20)    \\
\hline
$\beta_5$  &   4.54  (4.18,4.87) & 4.50  (4.31,4.67) &  4.57(4.20,4.92) &  4.50 (4.33,4.64)    \\
\hline
$\sigma_{\varepsilon}$  &  1.10  (0.87,2.64) &  0.50 (0.41,0.58) & 1.09 (0.87,2.47) & 0.50 (0.43,0.58)    \\
\hline
$\delta_{\varepsilon}$  &    -- &   1.91 (1.33,2.84) &  -- & 1.96  (1.52,2.86)    \\
\hline
$\sigma$  &  0.52  (0.26,1.52) & 0.48  (0.34,0.76) & 0.20 (0.06,0.40) & 0.28  (0.18,0.41)    \\
\hline
$\delta$  &    -- &   -- &  1.65 (0.97,5.07) &   2.17 (1.31,6.55)    \\
\hline
\end{tabular}
\caption{ Simulation study II with Scenario II: Posterior medians and 95\% credible intervals of the median estimators.}\label{table:SimulationII}
\end{center}
\end{table}

From Table \ref{table:SimulationII} we conclude that wrongly assuming normality of the distributions tends to make inference less precise, especially if we get the tails of the residual errors wrong. It also seems that the (more latent) tail parameter of the random effects is harder to estimate than that of the $\varepsilon_{ij}$'s.

\subsection{Framingham heart study}

We now illustrate the performance of the proposed models with real data. We use the data set reported in \cite{ZD01}, which consists of measurements of the cholesterol level for 200 randomly selected individuals from the Framingham heart study. The measurements were taken at the beginning of the study and then every 2 years for 10 years. We are interested in the relationship between the cholesterol level and the age (at baseline) and gender of the patients. \cite{ZD01} model this relationship through the LMM:
\begin{eqnarray}\label{FraminghamModel}
y_{ij} = \beta_1 +  \beta_2 \text{age}_i + \beta_3 \text{sex}_i + \beta_4 t_{ij} + u_{1i} + u_{2i} t_{ij} + \varepsilon_{ij},
\end{eqnarray}
\noindent where $y_{ij}$ represents the cholesterol level divided by 100 at the $j$th time for subject $i$ and $t_{ij}$ is $(\text{time}-5)/10$, with time measured in years from baseline. \cite{ZD01} assume normal residual errors $\varepsilon_{ij}$ and the density of the random effects ${\bf u}_i=(u_{1i}, u_{2i})^\top$ is represented by a semi-nonparametric truncated series expansion.

Model (\ref{FraminghamModel}) is simply the LMM in (\ref{LMM}) with $p=4$ and $q=2$. Here we adopt the following hierarchical structure:
\begin{eqnarray*}
\varepsilon_{ij}\mid \sigma_{\varepsilon},\delta_{\varepsilon}  &\stackrel{i.i.d.}{\sim}& \text{t}({0},\sigma_{\varepsilon} ,\delta_{\varepsilon}), \notag\\
{\bf u}_i\mid \bm{\theta_1},\bm{\theta_2},\rho &\sim& \text{GC}(F_1,F_2\mid \rho),
\end{eqnarray*}
where GC denotes a bivariate Gaussian copula with marginals $F_1$ and $F_2$, and correlation parameter $\rho$. For these marginal distributions we use a two-piece sinh-arcsinh distribution with the epsilon--skew parameterisation \citep{ROH14,RS14b} which contains a scale parameter $\sigma_i>0$, a skewness parameter $\gamma_i\in (-1,1)$, and a kurtosis parameter $\delta_i>0$, $i=1,2$. We denote this model as Model 1. We adopt the prior structure:
\begin{eqnarray}\label{PriorFramingham}
\pi(\bm{\beta},\sigma_{\varepsilon},\delta_{\varepsilon},\sigma_1,\sigma_2,\gamma_1,\gamma_2,\delta_1,\delta_2,\rho) &\sim& \dfrac{\pi(\delta_{\varepsilon})\pi(\rho)}{\sigma_{\varepsilon}} \prod_{i=1}^2 \pi(\sigma_i)\pi(\gamma_i)\pi(\delta_i).
\end{eqnarray}
 As a weakly informative prior for the degrees of freedom of the Student-$t$ distribution, $\delta_{\varepsilon}$ we employ the prior proposed in \cite{RS14b}. For each of the scale parameters $(\sigma_1,\sigma_2$) we adopt a half-Cauchy prior \citep{PS12}. The shape parameters $(\delta_1,\delta_2)$ are assigned the prior proposed in \cite{RS14b} for a general class of kurtosis parameters. For each of the skewness parameters $(\gamma_1,\gamma_2)$ we adopt a uniform distribution on $(-1,1)$. In a bivariate Gaussian copula the Spearman measure of association, $r_{\rho}\in(-1,1)$, can be calculated in closed form as
\begin{eqnarray*}
r_{\rho} = \dfrac{6}{\pi}\arcsin\left(\dfrac{\rho}{2}\right),
\end{eqnarray*}
for which we assume a uniform prior $r_{\rho}\sim U(-1,1)$, inducing the following prior density on $\rho$
\begin{eqnarray*}
\pi(\rho) \propto \dfrac{1}{\sqrt{1-\left({\rho}/{2}\right)^2}}.
\end{eqnarray*}

The propriety of the corresponding posterior distribution is guaranteed by Theorem \ref{FOSSMNPPExt1}. We also consider the following submodels: Model 2 $(\delta_1,\delta_2)=(1,1)$ (two--piece normal marginal random effects); Model 3 $(\delta_1,\delta_2,\gamma_2)=(1,1,0)$ (one marginally normal random effect and one marginally two-piece normal random effect); Model 4 $(\delta_1,\delta_2,\gamma_1,\gamma_2) = (1,1,0,0)$ (normal random effects); and Model 5 $(\delta_1,\delta_2,\gamma_1,\gamma_2,\delta_{\varepsilon}) = (1,1,0,0,\infty)$ (normal random effects and normal residual errors). Finally, Model 6 assumes normality of all distributions as well as independence between both random effects. A posterior sample of size $5,000$ was obtained after a burn-in period of $15,000$ iterations and thinning every $25$ iterations ($140,000$ simulations in total). Table \ref{table:SummaryPost} presents a summary of the posterior results.

The Bayes factors, calculated using the Savage-Dickey density ratio, slightly favour the general model. However, the evidence in favour of this model is not decisive (see Table \ref{table:BF}). The Bayes factors associated to Models 4--6 (against Model 1) are not shown in Table \ref{table:BF} since they are virtually zero, but the Bayes factors associated to other sub-models are presented in this table for illustration. Bayes factors clearly indicate support for skewness of $u_{1t}$ and dependence of both random effects, which leaves us with Models 1-3. An argument of parsimony could, then, lead to selecting Model 3. Notice that the intervals for the general model are more spread out (\emph{e.g.}~for $\beta_4$ in Table \ref{table:SummaryPost}). This indicates that the sample does not contain enough information to accurately estimate all the parameters of this model. The posterior probability of $\{\delta_{\varepsilon} > 10\}$ in Models 1 and 3 is approximately $0.005$ and for Model 2 is approximately $0.015$, 
which suggests that models with heavier tails than normal are favoured.

To assess the predictive performance of the different models we use the log pseudomarginal likelihood (LPML), which is defined as the sum of the log conditional predictive ordinate (CPO) statistic for each observation. The CPO for subject $i$ is defined as the predictive density of the vector of observations ${\bf y}_i$ given the rest of the observations. This is, $\text{CPO}_i = \pi({\bf y}_i\mid {\bf y}\backslash \{{\bf y}_i\})$, where the predictive density is evaluated at the vector of observations corresponding to the $i^{th}$ subject (see \citealp{B12}). These quantities are approximated using a harmonic mean estimator as described in \cite{B12}.  Table \ref{table:SummaryPost} shows the LPML for the different models, which favour Model 3. A larger sample would likely provide stronger evidence for model selection.

The analysis in \cite{ZD01} suggests departures from normality in the distribution of random intercepts, while the estimated random slope distribution can be approximated by a normal density. Our conclusions are in line with the conclusions of the parametric, semi-nonparametric and the nonparametric analyses in \cite{ZD01}, \cite{J08}, and \cite{J09}. However, the models proposed here also allow us to interpret the meaning of the parameters, separating those controlling the shape of the distribution from those controlling the dependence between the random effects.

\begin{landscape}
\begin{table}[h!]
\small
\begin{center}
\begin{tabular}[h]{|c|c|c|c|c|c|c|c|}
\hline
Model & $\gamma_1=0$  & $\gamma_2=0$ & $\delta_1=1$ & $\delta_2=1$ & $\rho = 0$& $(\delta_1,\delta_2)=(1,1)$ &$(\gamma_2,\delta_1,\delta_2)=(0,1,1)$  \\
\hline
BF & 0.04 & 0.65 & 0.77 & 1.04 & 0.02 & 0.63 & 0.42 \\
\hline
\end{tabular}
\caption{ Framingham data: Bayes factors (against the general Model 1) for several hypotheses.}\label{table:BF}
\end{center}
\end{table}

\begin{table}[h!]
\small
{
\begin{tabular}[h]{|c c c c c c c |}
\hline
Model & 1  & 2 & 3  & 4 & 5 & 6 \\
\hline
$\beta_1$ (intercept) & 1.494 (1.218,1.794) &  1.649 (1.436,1.973) & 1.640 (1.366,1.878) & 1.621 (1.336,1.957)  & 1.610 (1.223,1.939)  & 1.715 (1.399,1.980)  \\
\hline
$\beta_2$ (age) & 0.016 (0.008,0.024) &  0.012 (0.005,0.017) & 0.013 (0.007,0.020) & 0.017 (0.010,0.024)  & 0.018 (0.010,0.027) &  0.015 (0.009, 0.023)  \\
\hline
$\beta_3$  (sex) & -0.057 (-0.154,0.044) &   -0.059 (-0.162,0.040) &  -0.061 (-0.159,0.042) & -0.048 (-0.159,0.074)  & -0.060 (-0.172, 0.048)  &  -0.015 (-0.123,0.094) \\
\hline
$\beta_4$  (time) &  0.130 (-0.011,0.384) &  0.121 (-0.002,0.321) & 0.290 (0.243,0.336) &  0.289 (0.241,0.337)  & 0.282 (0.233,0.327)   & 0.282 (0.235,0.329) \\
\hline
$\sigma_{\varepsilon}$ & 0.163 (.146,0.182) &  0.163 (0.147,0.181) & 0.164 (0.147,0.182) &  0.163 (.146,0.182)  & 0.209 (0.198,0.221) &  0.209 (0.198,0.220)  \\
\hline
$\delta_{\varepsilon}$ &  5.094 (3.513,8.095) & 5.116  (3.522,8.153) &  5.202 (3.586,8.459) & 5.115 (3.524,8.235)  & --   &  -- \\
\hline
$\sigma_1$ &  0.275 (0.181,0.412) & 0.372  (0.336,0.418) & 0.372 (0.335,0.416) & 0.384 (0.347,0.428)  & 0.380 (0.344,0.425)  & 0.376 (0.340,0.419) \\
\hline
$\sigma_2$ & 0.201 (0.075,0.829) & 0.186 (0.133,0.244) & 0.203 (0.144,0.257) &  0.204 (0.147,0.261)  &  0.193 (0.125,0.254) & 0.193 (0.127,0.252)  \\
\hline
$\gamma_1$ &  -0.331 (-0.524,-0.115) & -0.345 (-0.521,-0.145) &  -0.339 (-0.524,-0.138) & -- & --  &  -- \\
\hline
$\gamma_2$ & -0.546 (-0.972,0.277)  &  -0.597 (-0.969,0.086)   &  --   & -- & --  &  --  \\
\hline
$\delta_1$ & 0.820 (0.656,1.057)  &  --  &   --  & -- &  --  & -- \\
\hline
$\delta_2$ & 1.047 (0.630,3.407)  &  -- &   --  & -- & --  &  --  \\
\hline
$\rho$ & 0.425 (0.177,0.640) &  0.397 (0.158,0.611) & 0.383 (0.137,0.620) & 0.397 (0.141,0.638)  & 0.4243 (0.168,0.686) &  --  \\
\hline
LPML & -22.81   & -25.24  &  -19.12  & -26.75 & -30.88 &  -32.98 \\
\hline
\end{tabular}
}
\caption{ Framingham data: Summary of the posterior samples (medians and 95\% credible intervals) and LPML.}\label{table:SummaryPost}
\end{table}
\end{landscape}


\subsection{HIV-1 viral load after unstructured treatment interruption}

Here we revisit the data set analysed in \cite{VL09}, which concerns the study of 72 perinatally HIV-infected children and adolescents. Some of the subjects present unstructured treatment interruption, which is a common phenomenon among HIV-infected people, due mainly to treatment fatigue \citep{VL09}. The number of observations from baseline
(month 0) to month 24 ranges from 13 to 71. Out of 362 observations, 26 observations (7\%) were below the detection limits and were censored at these values. \cite{VL09} proposed the model:
\begin{eqnarray*}
y_{ij} = \beta_j + u_i + \varepsilon_{ij},
\end{eqnarray*}
where $y_{ij}$ is the $\log_{10}$ HIV RNA (HIV Ribonucleic acid, which is used to monitor the status of HIV) for subject $i$ at time $t_j$, using $t_1=0$, $t_2=1$, $t_3=3$, $t_4=6$, $t_5=9$, $t_6=12$, $t_7=18$ and $t_8=24$ months.

We allow for Student-$t$ tails in the random effects and the errors (with zero locations), and thus use Model 4 in Subsection \ref{Sec:SimII} with the same prior structure as in (\ref{PriorSimulationII}). We also try the simpler models 1-3 as defined in Subsection \ref{Sec:SimII}.

The propriety of the posterior is ensured by Theorem \ref{CenUncMEAFT}.

 Table \ref{table:SummaryPost2} reveals clear evidence for fat tails in both distributions, especially the residual errors. The predictive criterion LPML also favours the general model and particularly penalizes model with normal residual errors.
Clearly, ignoring the heavy tails affects the inference on the fixed effects, especially for $\beta_1$.

\begin{table}[h!]
\small
\begin{center}
{
\begin{tabular}[h]{|c c c c c|}
\hline
Model                 & 1                  & 2                 & 3                   & 4                  \\
\hline
$\beta_1$             &  3.62 (3.36,3.87)  & 4.00 (3.76,4.24)  & 3.70 (3.45,3.93)    & 4.16 (3.95,4.38)    \\
\hline
$\beta_2$             &  4.19 (3.93,4.43)  & 4.22 (4.00,4.45)  & 4.26 (4.02,4.50)    & 4.37 (4.18,4.56)     \\
\hline
$\beta_3$             &  4.26 (4.00,4.52)  & 4.26 (4.04,4.48)  & 4.34 (4.09,4.59)    & 4.41 (4.22,4.60)     \\
\hline
$\beta_4$             &  4.38 (4.12,4.64)  & 4.43 (4.20,4.65)  & 4.46 (4.21,4.70)    & 4.58 (4.38,4.77)      \\
\hline
$\beta_5$             &  4.61 (4.33,4.89)  & 4.53 (4.31,4.76)  & 4.68 (4.41,4.96)    & 4.68 (4.48,4.87)      \\
\hline
$\beta_6$             &  4.60 (4.30,4.89)  & 4.51 (4.28,4.74)  & 4.67 (4.38,4.95)    & 4.65 (4.45,4.85)      \\
\hline
$\beta_7$             &  4.70 (4.36,5.02)  & 4.70 (4.45,4.95)  & 4.77 (4.46,5.08)    & 4.84 (4.61,5.06)     \\
\hline
$\beta_8$             &  4.81 (4.40,5.22)  & 4.73 (4.44,5.03)  & 4.88 (4.47,5.28)    & 4.88 (4.62,5.14)      \\
\hline
$\sigma_{\varepsilon}$&  0.60 (0.55,0.66)  &  0.20 (0.16,0.25) & 0.61 (0.56,0.66)    & 0.20 (0.16,0.24)      \\
\hline
$\delta_{\varepsilon}$& --                 & 1.27  (0.97,1.66) & --                  & 1.30 (1.01,1.69)      \\
\hline
$\sigma$              & 0.90 (0.75,1.10)   & 0.88  (0.72,1.07) & 0.66 (0.46,0.90)    & 0.56 (0.37,0.79)    \\
\hline
$\delta$              & --                 &   --              & 3.86 ( 1.65,16.81)  & 2.48 (1.24,7.71)    \\
\hline
LPML                  &  -397.4            & -291.4            &  -401.5             & -287.6                \\
\hline
\end{tabular}
}
\caption{ HIV data: Summary of the posterior samples (sample median and 95\% credible intervals) and LPML.}\label{table:SummaryPost2}
\end{center}
\end{table}

Finally, the predictive performance as measured by LPML is substantially better for our models 4 and 2 than for the best model in \cite{B12}, who obtain $\text{LPML}=-366.27$ on the same data using the Skew-contaminated normal model of \cite{L10}, which has normal tails for both random effects and error distributions. Also, their posterior distribution of $\beta_1$ is substantially shifted towards smaller values (consistent with our models 1 and 3, which impose normal errors).
\section{Discussion}

This paper focuses on hierarchical linear mixed models which are used frequently for both longitudinal and survival modelling in a number of application areas, such as medicine, epidemiology and economics. These models are often based on simple distributional assumptions, such as normality for both the residual errors and the random effects. There is ample evidence in the literature that such assumptions can seriously affect inference on the random effects and the predictive distribution (see \textit{e.g.}~\citealp{LT08,ZD01}) and we add to that evidence in our analysis of simulated data. Misspecification of the distribution of the residual errors and the random effects tends to have a substantial effect on the predictive distribution, which explicitly depends on the aforementioned distributions. The use of normal assumptions when the true generating model is not normal also has a marked effect on the estimation of the variance of the residual errors and the random effects, as well as a decrease in the precision of the estimation of the fixed effects.

We consider Bayesian inference for these models with flexible distributions and sensible prior structures. These priors are intended to convey a lack of strong prior information; as they are based on a combination of formal arguments (such as invariance) and pragmatic common sense, they end up being improper. Our results provide a formal justification for Bayesian inference in these wide classes of models: our models accommodate any proper distribution for the random effects (which could even be dependent, if required) combined with two-piece scale mixtures of normal distributions for the residual errors, as well as potential censoring. Our results thus cover all random effects distributions (both parametric and nonparametric) that were proposed in the literature except for those that are dependent on the residual errors (such as those in \citealp{L10}, which where used in \citealp{B12}). The conditions for the propriety of the posterior distribution are rather mild and easy to check in practice, especially compared to those in previous papers \citep{HC96,S01,R14}, which obtain sets of conditions that combine the sample size, the number of clusters, the design matrix, and the value of the hyperparameters of the priors on the random effects. This also suggests that most of the posterior impropriety issues come from assuming improper priors on the ``deeper'' parameters of the random effects.

The analysis of simulated datasets leads to quite promising results, which suggest that priors are well-calibrated and sensible, and Bayes factors seem to provide reliable indications. In addition, Appendix C in the Supplementary Material presents results for the case where both the residual errors and the random effects display skewness, which corroborate the results presented in the main text.

Inference on both real datasets strongly indicates departures from normality in that the residual error distributions have much fatter tails than normal. In addition, the Framingham heart data support dependence between the random effect components with normal random slope and skew-normal random intercept. The HIV data present clear evidence for heavy-tailed random effects.
By using simple Student-$t$ distributions on both errors and random effects, we obtain considerably better predictive results on the HIV data than those presented by \cite {B12}, using their preferred model. We remind the reader that the distributions considered in \cite{B12} are multivariate scale mixtures of skew-normal distributions, so perhaps might be affected by the kind of issues we describe for multivariate scale mixtures of normals in Section \ref{WarningCase}. For the real data, evidence based on Bayes factors points in the same direction as the predictive performance measured by LPML, which is reassuring.

Here we have employed the R package `spBayes' in all of our numerical examples, mainly for reasons of efficiency and simplicity. However, the user can also implement the models proposed here in any other ``general purpose'' MCMC software such as Stan (gradient-based MCMC), PROC MCMC (from SAS, which implements a random walk Metropolis), or any package which includes a Metropolis-within-Gibbs sampler.

\section*{Supplementary material}

Supplementary material includes proofs not mentioned in the text (Appendix A), conditions on the mixing distributions (Appendix B) and a further simulation study with skewed errors and random effects (Appendix C). The R codes used in the real data examples are available at \url{http://www.rpubs.com/FJRubio}.

\section*{Acknowledgements}
We thank two referees, an Associate Editor, and the Editor for helpful comments.

\bibliographystyle{plainnat}
\bibliography{references}


\section*{SUPPLEMENTARY MATERIAL}

\section*{Appendix A: Proofs}

\subsection*{Proof of Remark \ref{FOSSMN}}
The marginal density of the data is given by
\begin{eqnarray*}
\tilde{m}({\bf y}) &=& \int p({\bf y}\mid {\bf u}, \bm{\beta}, \sigma_{\varepsilon},\delta_{\varepsilon}) p({\bf u}) \pi(\bm{\beta}) \dfrac{1}{\sigma_{\varepsilon}^{b+1}} \pi(\delta_{\varepsilon}) \, d{\bf u} d\bm{\beta} d\sigma_{\varepsilon} d\delta_{\varepsilon}\\
&=& \int \dfrac{\tau^{n/2}}{\sigma_{\varepsilon}^n} \prod_{i=1}^r\prod_{j=1}^{n_i} \phi \left[\dfrac{\tau^{\frac{1}{2}}}{\sigma_{\varepsilon}}(y_{ij}-{\bf x}_{ij}^{\top}\bm{\beta} - {\bf z}_{ij}^{\top}{\bf u}_i)\right] \\
&\times& p({\bf u}) \pi(\bm{\beta}) \dfrac{1}{\sigma_{\varepsilon}^{b+1}} \pi(\delta_{\varepsilon})
 d{\bf u} d\bm{\beta} d\sigma_{\varepsilon} d H_{\varepsilon}(\tau\mid\delta_{\varepsilon}) d\delta_{\varepsilon} .
\end{eqnarray*}

Consider the change of variable $\tilde{\sigma}_{\varepsilon} = \sigma_{\varepsilon}/{\tau}^{\frac{1}{2}}$,  with corresponding Jacobian $\vert {\mathcal J} \vert = {\tau}^{\frac{1}{2}}$. Then,
\begin{eqnarray*}
\tilde{m}({\bf y}) &=& \int \dfrac{1}{\tilde{\sigma}_{\varepsilon}^n} \prod_{i=1}^r\prod_{j=1}^{n_i} \phi\left[\dfrac{1}{\tilde{\sigma}_{\varepsilon}}(y_{ij}-{\bf x}_{ij}^{\top}\bm{\beta} - {\bf z}_{ij}^{\top}{\bf u}_i)\right] \notag\\
&\times&p({\bf u}) \dfrac{1}{\tilde{\sigma}_{\varepsilon}^{b+1}}  \pi(\bm{\beta}) \pi(\delta_{\varepsilon}) {\tau}^{-\frac{b}{2}} d{\bf u} d\bm{\beta} d\tilde{\sigma}_{\varepsilon} d H_{\varepsilon}({\tau}\mid \delta_{\varepsilon}) d\delta_{\varepsilon} \notag\\
&=& m({\bf y})\int_{\Delta}\int_{{\mathbb R}_+} {\tau}^{-\frac{b}{2}} \pi(\delta_{\varepsilon})d H_{\varepsilon}({\tau}\mid\delta_{\varepsilon}) d\delta_{\varepsilon},
\end{eqnarray*}
where we can identify the first factor as the marginal density of the data associated to the model with normal distributional assumptions on $\bm{\varepsilon}$.

\subsection*{Proof of Theorem \ref{FOSSMNPPExt1}}

For practical purposes we will work with $\sigma_{\varepsilon}^2$ instead of $\sigma_{\varepsilon}$. The marginal density of the data is given by
\begin{eqnarray}\label{marginalModelII}
\tilde{m}({\bf y}) &=& \int p({\bf y}\mid {\bf u}, \bm{\beta}, \sigma_{\varepsilon}^2,\delta_{\varepsilon}) p({\bf u}) \pi(\bm{\beta}) \dfrac{1}{(\sigma_{\varepsilon}^2)^{\frac{b}{2}+1}} \pi(\delta_{\varepsilon}) \, d{\bf u} d\bm{\beta} d\sigma_{\varepsilon}^2 d\delta_{\varepsilon}\notag\\
&=& \int  \prod_{i=1}^r\prod_{j=1}^{n_i}\dfrac{\tau_{ij}^{\frac{1}{2}}}{\sqrt{2\pi \sigma_{\varepsilon}^2}} \exp\left[-\dfrac{\tau_{ij}}{2\sigma_{\varepsilon}^2}(y_{ij}-{\bf x}_{ij}^{\top}\bm{\beta} - {\bf z}_{ij}^{\top}{\bf u}_i)^2\right] \notag\\
&\times& p({\bf u}) \pi(\bm{\beta}) \dfrac{1}{(\sigma_{\varepsilon}^2)^{\frac{b}{2}+1}} \pi(\delta_{\varepsilon})
 d{\bf u} d\bm{\beta} d\sigma_{\varepsilon}^2 \left[ \prod_{i=1}^r\prod_{j=1}^{n_i} d H_{\varepsilon}(\tau_{ij}\mid\delta_{\varepsilon}) \right] d\delta_{\varepsilon} .
\end{eqnarray}

Denote $\tau_{+} = \max_{i,j}\tau_{ij}$ and $\tau_{-} = \min_{i,j}\tau_{ij}$. Then, we can obtain the following lower bound
\begin{eqnarray*}
\tilde{m}({\bf y}) &\geq& \int \prod_{i=1}^r\prod_{j=1}^{n_i}\dfrac{\tau_{-}^{\frac{1}{2}}}{\sqrt{2\pi \sigma_{\varepsilon}^2}} \exp\left[-\dfrac{\tau_{+}}{2\sigma_{\varepsilon}^2}(y_{ij}-{\bf x}_{ij}^{\top}\bm{\beta} - {\bf z}_{ij}^{\top}{\bf u}_i)^2\right] \\
&\times& p({\bf u}) \pi(\bm{\beta}) \dfrac{1}{(\sigma_{\varepsilon}^2)^{\frac{b}{2}+1}} \pi(\delta_{\varepsilon})
 d{\bf u} d\bm{\beta} d\sigma_{\varepsilon}^2 \left[\prod_{i=1}^r\prod_{j=1}^{n_i} d H_{\varepsilon}(\tau_{ij}\mid\delta_{\varepsilon}) \right] d\delta_{\varepsilon} .
\end{eqnarray*}

Consider the change of variable $\tilde{\sigma}_{\varepsilon}^2 = \dfrac{\sigma_{\varepsilon}^2}{\tau_{+}}$
\begin{eqnarray*}
\tilde{m}({\bf y}) &\geq& \int \prod_{i=1}^r\prod_{j=1}^{n_i}\dfrac{\tau_{-}^{\frac{1}{2}}}{\sqrt{2\pi \tilde{\sigma}_{\varepsilon}^2\tau_{+}}} \exp\left[-\dfrac{1}{2\tilde{\sigma}_{\varepsilon}^2}(y_{ij}-{\bf x}_{ij}^{\top}\bm{\beta} - {\bf z}_{ij}^{\top}{\bf u}_i)^2\right] \\
&\times& \tau_{+} p({\bf u}) \pi(\bm{\beta}) \dfrac{1}{(\sigma_{\varepsilon}^2\tau_{+})^{\frac{b}{2}+1}} \pi(\delta_{\varepsilon})
 d{\bf u} d\bm{\beta} d\sigma_{\varepsilon}^2 \left[\prod_{i=1}^r\prod_{j=1}^{n_i} d H_{\varepsilon}(\tau_{ij}\mid\delta_{\varepsilon}) \right] d\delta_{\varepsilon} \\
 &=& m({\bf y}) \int \left(\dfrac{\tau_{-}^{n/2}}{\tau_{+}^{n/2}}\right) \tau_{+}^{-\frac{b}{2}} \pi(\delta_{\varepsilon})\prod_{i=1}^r\prod_{j=1}^{n_i} d H_{\varepsilon}(\tau_{ij}\mid\delta_{\varepsilon}) d\delta_{\varepsilon},
\end{eqnarray*}
Using the latter expression together with Theorem 1 from \cite{FOS97} it follows that (a) is a necessary condition for the propriety of the posterior.

Now, define ${\mathcal T}= \operatorname{diag}(\tau_1,\dots,\tau_n)$ (the diagonal matrix of mixing variables), ${\tilde {\bf y}}={\mathcal T}^{\frac{1}{2}}{\bf y}$, ${\tilde {\bf X}}={\mathcal T}^{\frac{1}{2}}{\bf X}$, and ${\tilde {\bf Z}}={\mathcal T}^{\frac{1}{2}}{\bf Z}$, where ${\mathcal T}^{\frac{1}{2}}= \operatorname{diag}(\sqrt{\tau_1},\dots,\sqrt{\tau_n})$. Then, by using that $\pi(\bm{\beta})\leq K$, we can obtain the following upper bound for (\ref{marginalModelII})
\begin{eqnarray*}
\tilde{m}({\bf y}) &\leq& K\int \dfrac{\det({\mathcal T})^{\frac{1}{2}}}{(2\pi \sigma_{\varepsilon}^2)^{\frac{n}{2}}} \exp\left[-\dfrac{1}{2\sigma_{\varepsilon}^2}({\tilde {\bf y}} - {\tilde {\bf X}}\bm{\beta} - {\tilde {\bf Z}}{\bf u})^{\top} ({\tilde {\bf y}} - {\tilde {\bf X}}\bm{\beta} - {\tilde {\bf Z}}{\bf u})\right]\\
&\times& p({\bf u})  \dfrac{1}{(\sigma_{\varepsilon}^2)^{\frac{b}{2}+1}} \pi(\delta_{\varepsilon})
 d{\bf u} d\bm{\beta} d\sigma_{\varepsilon}^2 \left[\prod_{i=1}^r\prod_{j=1}^{n_i} d H_{\varepsilon}(\tau_{ij}\mid \delta_{\varepsilon}) \right] d\delta_{\varepsilon} .
\end{eqnarray*}

Following the proof of Theorem 1 in \cite{FOS97} we have
\begin{eqnarray*}
({\tilde {\bf y}} - {\tilde {\bf X}}\bm{\beta} - {\tilde {\bf Z}}{\bf u})^{\top} ({\tilde {\bf y}} - {\tilde {\bf X}}\bm{\beta} - {\tilde {\bf Z}}{\bf u}) = (\bm{\beta}-\hat{\bm{\beta}})^{\top}{\tilde {\bf X}}^{\top}{\tilde {\bf X}}(\bm{\beta}-\hat{\bm{\beta}}) + c({\tilde{\bf Z}},{\tilde{\bf y}}),
\end{eqnarray*}
where $\hat{\bm{\beta}}= ({\tilde {\bf X}}^{\top}{\tilde {\bf X}})^{-1}{\tilde {\bf X}}^{\top}({\tilde {\bf y}} - {\tilde {\bf Z}}{\bf u})$ and $c({\tilde{\bf Z}},{\tilde{\bf y}}) = ({\tilde {\bf y}} - {\tilde {\bf Z}}{\bf u})^{\top} {\bf M}_{{\tilde{\bf X}}}({\tilde {\bf y}} - {\tilde {\bf Z}}{\bf u})$, with ${\bf M}_{{\tilde{\bf X}}} = {\bf I}_n - {\tilde {\bf X}}({\tilde {\bf X}}^{\top}{\tilde {\bf X}})^{-1}{\tilde {\bf X}}^{\top}$. Then, we get
\begin{eqnarray*}
\tilde{m}({\bf y}) &\leq& K \int \dfrac{\det({\mathcal T})^{\frac{1}{2}}}{(2\pi \sigma_{\varepsilon}^2)^{\frac{n}{2}}} \exp\left[-\dfrac{1}{2\sigma_{\varepsilon}^2}(\bm{\beta}-\hat{\bm{\beta}})^{\top}{\tilde {\bf X}}^{\top}{\tilde {\bf X}}(\bm{\beta}-\hat{\bm{\beta}})\right] \exp\left[-\dfrac{c({\tilde{\bf Z}},{\tilde{\bf y}})}{2\sigma_{\varepsilon}^2}\right]\\
&\times& p({\bf u})  \dfrac{1}{(\sigma_{\varepsilon}^2)^{\frac{b}{2}+1}} \pi(\delta_{\varepsilon})
 d{\bf u} d\bm{\beta} d\sigma_{\varepsilon}^2 \left[\prod_{i=1}^r\prod_{j=1}^{n_i} d H_{\varepsilon}(\tau_{ij}\mid\delta_{\varepsilon}) \right] d\delta_{\varepsilon} .
\end{eqnarray*}

By integrating $\bm{\beta}$ out we obtain
\begin{eqnarray*}
\tilde{m}({\bf y}) &\leq& K\int \dfrac{\det({\mathcal T})^{\frac{1}{2}}}{\det({\tilde {\bf X}}^{\top}{\tilde {\bf X}})^{\frac{1}{2}}(2\pi \sigma_{\varepsilon}^2)^{\frac{n-p}{2}}} \exp\left[-\dfrac{c({\tilde{\bf Z}},{\tilde{\bf y}})}{2\sigma_{\varepsilon}^2}\right]\\
&\times& p({\bf u})  \dfrac{1}{(\sigma_{\varepsilon}^2)^{\frac{b}{2}+1}} \pi(\delta_{\varepsilon})
 d{\bf u} d\sigma_{\varepsilon}^2 \left[\prod_{i=1}^r\prod_{j=1}^{n_i} d H_{\varepsilon}(\tau_{ij}\mid\delta_{\varepsilon}) \right] d\delta_{\varepsilon} .
\end{eqnarray*}

Now, after integrating $\sigma_{\varepsilon}^2$ we obtain
\begin{eqnarray}\label{margineq1}
\tilde{m}({\bf y}) &\leq& K_1\int \dfrac{\det({\mathcal T})^{\frac{1}{2}}}{\det({\tilde {\bf X}}^{\top}{\tilde {\bf X}})^{\frac{1}{2}}} c({\tilde{\bf Z}},{\tilde{\bf y}})^{-\frac{n+b-p}{2}}  p({\bf u})\pi(\delta_{\varepsilon})
 d{\bf u} \left[\prod_{i=1}^r\prod_{j=1}^{n_i} d H_{\varepsilon}(\tau_{ij}\mid\delta_{\varepsilon}) \right] d\delta_{\varepsilon},
\end{eqnarray}
where $K_1>0$ is a known constant. Note that $\det({\tilde {\bf X}}^{\top}{\tilde {\bf X}}) = \det({ {\bf X}}^{\top}{\mathcal T}{ {\bf X}})$, and
\begin{eqnarray*}
c({\tilde{\bf Z}},{\tilde{\bf y}}) &=&({ {\bf y}} - { {\bf Z}}{\bf u})^{\top} {\mathcal T}^{\frac{1}{2}} {\bf M}_{{\tilde{\bf X}}}{\mathcal T}^{\frac{1}{2}} ({ {\bf y}} - { {\bf Z}}{\bf u})\\
&=& ({ {\bf y}} - { {\bf Z}}{\bf u})^{\top} {\mathcal T}({ {\bf y}} - { {\bf Z}}{\bf u}) + ({ {\bf y}} - { {\bf Z}}{\bf u})^{\top}{\mathcal T}{ {\bf X}}({ {\bf X}}^{\top}{\mathcal T}{{\bf X}})^{-1}{{\bf X}}^{\top}{\mathcal T}({ {\bf y}} - { {\bf Z}}{\bf u}).
\end{eqnarray*}


From Lemma 1 from \cite{FS00} we know that $\det({ {\bf X}}^{\top}{\mathcal T}{ {\bf X}})$ has upper and lower bounds that are both proportional to $\prod_{i=1}^p \tau_{m_i} =\max\{\prod_{i=1}^p \tau_{s_i}: 1\leq s_1 <\dots< s_p\leq n\text{ and } \det({\bf x}_{s_1},\dots,{\bf x}_{s_p})\neq 0\}$, where $\det({\bf x}_{s_1},\dots,{\bf x}_{s_p})$ denotes the determinant of the submatrix of $X$ corresponding to the observations $y_{s_1},\dots, y_{s_p}$. Define also $\tau_c = \max\{\tau_i: i\neq m_1, \dots, m_p\}$. Using these results we obtain the following inequality
\begin{eqnarray*}
\dfrac{\det({\mathcal T})^{\frac{1}{2}}}{\det({\tilde {\bf X}}^{\top}{\tilde {\bf X}})^{\frac{1}{2}}} c({\tilde{\bf Z}},{\tilde{\bf y}})^{-\frac{n+b-p}{2}} &\leq& K_2 \dfrac{\prod_{j\neq m_1,\dots,m_p} \tau_j^{\frac{1}{2}} }{c({\tilde{\bf Z}},{\tilde{\bf y}})^{\frac{n+b-p}{2}}}\\
 &\leq& K_2 \dfrac{\tau_c^{\frac{n-p}{2}}}{c({\tilde{\bf Z}},{\tilde{\bf y}})^{\frac{n+b-p}{2}} },
\end{eqnarray*}
where $K_2>0$ is a constant. Now, by defining the $n\times (p+1)$ matrix ${\bf L} = ({\bf X}:{\bf y}-{\bf Z}{\bf u})$ and subsequently applying the Binet--Cauchy theorem \citep{FS00}, we obtain
\begin{eqnarray*}
c({\tilde{\bf Z}},{\tilde{\bf y}}) &=& \dfrac{\det({\bf L}^{\top} {\mathcal T} {\bf L})}{\det({\bf X}^{\top} {\mathcal T} {\bf X})} \\
 &=& \dfrac{1}{\det({\bf X}^{\top} {\mathcal T} {\bf X})}  \sum_{1\leq s_1 <\dots < s_{p+1}\leq n} \left(\prod_{j=1}^{p+1} \tau_{s_j}\right) {\det} ^2 \left( \begin{array}{c c c} {\bf x}_{s_1} & \dots & {\bf x}_{s_{p+1}} \\ y_{s_1} - {\bf z}_{s_1}^{\top}{\bf u} & \dots & y_{s_{p+1}} - {\bf z}_{s_{p+1}}^{\top}{\bf u}  \end{array} \right).
\end{eqnarray*}

Given that ${\bf y}-{\bf Z}{\bf u}$ is not in the column space of ${\bf X}$ we have that $c({\tilde{\bf Z}},{\tilde{\bf y}})>0$. Consequently
$$c({\tilde{\bf Z}},{\tilde{\bf y}}) \geq P({\bf u}) \tau_c,$$
 with
$$P({\bf u})=\min_{1\leq s_1 <\dots < s_{p+1}\leq n} {\det}^2 \left( \begin{array}{c c c} {\bf x}_{s_1} & \dots & {\bf x}_{s_{p+1}} \\ y_{s_1} - {\bf z}_{s_1}^{\top}{\bf u} & \dots & y_{s_{p+1}} - {\bf z}_{s_{p+1}}^{\top}{\bf u}  \end{array} \right),$$
where the minimum is taken over the non--zero determinants. Then,
\begin{eqnarray*}
\dfrac{\det({\mathcal T})^{\frac{1}{2}}}{\det({\tilde {\bf X}}^{\top}{\tilde {\bf X}})^{\frac{1}{2}}} c({\tilde{\bf Z}},{\tilde{\bf y}})^{-\frac{n+b-p}{2}}
 &\leq& K_2 \dfrac{\tau_c^{-\frac{b}{2}}}{P({\bf u})^{\frac{n+b-p}{2}} }.
\end{eqnarray*}

Now, note that $P({\bf u})$, being a quadratic polynomial in $q$ variables, is a continuous coercive function (see Definition 2.4 in \citealp{G10}). Then, by Corollary 2.5 in \cite{G10}, it follows that $P({\bf u})$ achieves a global minimum. Given that $P({\bf u})>0$ by assumption, it follows that there exists $M>0$ such that $P({\bf u})>M$ for all ${\bf u}$. From this lower bound 
follows that 
\begin{eqnarray}\label{IneqCTau}
\dfrac{\det({\mathcal T})^{\frac{1}{2}}}{\det({\tilde {\bf X}}^{\top}{\tilde {\bf X}})^{\frac{1}{2}}} c({\tilde{\bf Z}},{\tilde{\bf y}})^{-\frac{n+b-p}{2}} &\leq& K_3 \tau_c^{-\frac{b}{2}},
\end{eqnarray}
for some constant $K_3>0$. Using this inequality in (\ref{margineq1}) and splitting the integral into the possible orderings of $\{\tau_1,\dots,\tau_n\}$ we can identify $\tau_c$ and the result follows.

\subsection*{Proof of Theorem \ref{CenUncMEAFT}}

The result follows by using that the censored observations contribute to the likelihood as factors in $[0,1]$. Then, we can obtain an upper bound for the marginal likelihood of the data which corresponds, up to a proportionality constant, to the marginal likelihood associated to the uncensored observations. The result then follows from Theorem \ref{FOSSMNPPExt1}.

\subsection*{Proof of Theorem \ref{CenMEAFT}}

By condition (d'), it follows that inequality (\ref{IneqCTau}) is satisfied for each ${\bf y} = {\bf y}^{\star}$, where ${\bf y}^{\star}\in {\mathcal E}$. Thus, the marginal likelihood of the data associated to each ${\bf y}^{\star}$ is finite (Theorem \ref{FOSSMNPPExt1}). Given that ${\mathcal E}$ is compact, we can obtain a finite upper bound for the marginal likelihood of the data by integrating out ${\bf y}^\star$ over ${\mathcal E}$, and the result follows.

\subsection*{Proof of Theorem \ref{FOSSMNPPExtAsym}}

By using the
 inequality
\begin{eqnarray*}
s(y\mid \mu,\sigma,\gamma,\delta) \leq \dfrac{2 M(\gamma)}{\sigma[a(\gamma)+b(\gamma)]}f\left(\dfrac{y-\mu}{\sigma M(\gamma)}\, \Big  \vert \, \delta\right),
\end{eqnarray*}
and the change of variable $\tilde{\sigma}_{\varepsilon}=\sigma_{\varepsilon} M(\gamma_{\varepsilon})$, we can obtain an upper bound for the marginal likelihood of the data which is a product of the marginal likelihood of model (\ref{LMM}), (\ref{Hierarchy2}), (\ref{scalarSMN}) with prior (\ref{Prior1}) and the expression in condition (e). Then, the result follows by using the proof of Theorem \ref{FOSSMNPPExt1}.


\section*{Appendix B: Conditions on the mixing distributions}

Note that the conditions on the mixing distributions in Corollary \ref{FOSSMNPP} and Theorem \ref{FOSSMNPPExt1} are related to the existence of their marginal moments of negative order.

\subsubsection*{Gamma mixing distribution}

Consider the mixing distribution $\tau\sim\text{Gamma}\left(\dfrac{\delta}{2},\dfrac{\delta}{2}\right)$ with density $f_{\Gamma}$. This mixing distribution produces a Student-$t$ sampling model. Let $c\geq 0$ and $\delta>2c$, then we have that

\begin{eqnarray*}
I=\int_0^{\infty}\int_0^{\infty} \tau^{-c} f_{\Gamma}(\tau\mid \delta) \pi(\delta)  d\tau d\delta = \int_0^{\infty}\dfrac{2^c \delta^{-c} \Gamma\left(\dfrac{\delta}{2}-c\right)}{\Gamma\left(\dfrac{\delta}{2}\right)}\pi(\delta) d\delta
\end{eqnarray*}

If $c=0$, it follows that $I<\infty$ given that $\pi(\delta)$ is proper. For $c>0$, it is necessary to use a proper prior $\pi(\delta)$ with support on $(2c+\epsilon,\infty)$, for any $\epsilon>0$.

\subsubsection*{Beta mixing distribution}

Consider the mixing distribution $\tau\sim\text{Beta}\left(\dfrac{\delta}{2},1\right)$ with density $f_{\beta}$. Let $c\geq0$ , then

\begin{eqnarray*}
I=\int_0^{\infty}\int_0^{\infty} \tau^{-c} f_{\beta}(\tau\mid \delta) \pi(\delta)  d\tau d\delta &=& \int_0^{\infty} \dfrac{\delta}{\delta - 2c} \pi(\delta) d\delta.
\end{eqnarray*}

For $c=0$, $I<\infty$ given that $\pi(\delta)$ is proper. However, for $c>0$ it is necessary to employ a truncated prior on $\delta$ with support on $(2c+\epsilon,\infty)$ for any $\epsilon>0$.

\subsubsection*{Birnbaum-Saunders mixing distribution}

Consider the mixing distribution $\tau\sim\text{BS}\left(\delta,\delta\right)$ with density $f_{BS}$ \citep{BS69}. Let $c\geq0$

\begin{eqnarray*}
I=\int_0^{\infty}\int_0^{\infty} \tau^{-c} f_{BS}(\tau\mid \delta) \pi(\delta)  d\tau d\delta &=& \int_0^{\infty} \frac{e^{\frac{1}{\delta ^2}} \delta ^{c-1}
   \left(K_{c-\frac{1}{2}}\left(\frac{1}{\delta
   ^2}\right)+K_{c+\frac{1}{2}}\left(\frac{1}{\delta
   ^2}\right)\right)}{\sqrt{2 \pi }} \pi(\delta) d\delta.
\end{eqnarray*}

\noindent where $K_n(z)$ represents the modified Bessel function of the second kind. For $c=0$ it follows that $I<\infty$, however, for $c>0$ this condition may rule out the use of certain heavy-tailed priors.

\section*{Appendix C: Simulation study III (skewed errors and random effects)}

In this section we study the model \eqref{LMMSimulation}. Four simulation scenarios are considered for the distribution of the residual errors $\epsilon_{ij}$ and the random effects $u_i$. For the first scenario we simulate from $\epsilon_{ij}\sim \text{N}(0,0.5)$ and $u_i\sim \text{N}(-1.5,0.5)$. For the second scenario we use $\epsilon_{ij}\sim \text{TPN}(0,0.5,0.5)$ and $u_i\sim \text{N}(-1.5,0.5)$. The third scenario consists of $\epsilon_{ij}\sim \text{N}(0,0.5)$ and $u_i\sim \text{TPN}(-1.5,0.5,0.5)$. The fourth scenario uses $\epsilon_{ij}\sim \text{TPN}(0,0.5,0.5)$ and $u_i\sim \text{TPN}(-1.5,0.5,0.5)$. We simulate $100$ data sets under these configurations. For each of these simulated samples, the model (\ref{LMMSimulation}) was fitted assuming that $\epsilon_{ij}\sim \text{TPN}(0,\sigma_{\varepsilon},\gamma_{\varepsilon})$ and $u_i \sim \text{TPN}(\mu,\sigma,\gamma)$ with the prior structure:
\begin{eqnarray}\label{PriorSimulation}
\pi(\beta_1,\beta_2,\sigma_{\varepsilon},\gamma_{\varepsilon},\mu,\sigma,\gamma) \propto \dfrac{\pi(\gamma_{\varepsilon})\pi(\mu)\pi(\sigma)\pi(\gamma)}{\sigma_{\varepsilon}},
\end{eqnarray}
\noindent where $\pi(\mu)$ is a uniform prior on $[-100,100]$, $\pi(\sigma)$ is a half-Cauchy density with mode $0$ and unit scale parameter and $\pi(\gamma)$ and $\pi(\gamma_{\varepsilon})$ are uniform priors on $(-1,1)$. The propriety of the posterior is then guaranteed by Theorem \ref{FOSSMNPPExt1}. For each of the $100$ simulated datasets we obtain a posterior sample of size $1,000$ 
after a burn-in period of $7500$ iterations and thinned every $10$ iterations (a total of $17,500$ MCMC draws). Table \ref{table:SimulationTP} summarizes the posterior samples and presents the median Bayes factors in favour of $\gamma_{\varepsilon}=0$ or $\gamma=0$ (symmetric errors or random effects), calculated using the Savage-Dickey density ratio. From this table, we can observe that despite the relatively small sample size, these Bayesian model selection criteria correctly identify the true model in each scenario.

In order to assess the impact of incorrectly imposing normality, we implement the model with normal residual errors and random effects. Table \ref{table:SimulationNTP} presents  posterior results for this model. The misspecification of the distribution of the residual errors and the random effects clearly affects the estimation of $\mu$ (which can be interpreted as an intercept).

\begin{table}[h!]
\begin{center}
\begin{tabular}[h]{|c|c|c|c|c|}
\hline
 & Scenario 1  & Scenario 2 & Scenario 3 & Scenario 4  \\
Parameter &  $\varepsilon_{ij}\sim \mbox{N}$ & $\varepsilon_{ij}\sim \mbox{TPN}$ & $\varepsilon_{ij}\sim \mbox{N}$  & $\varepsilon_{ij}\sim \mbox{TPN}$  \\
 & $u_i\sim \mbox{N}$  & $u_i\sim \mbox{N}$  & $u_i\sim \mbox{TPN}$  & $u_i\sim \mbox{TPN}$  \\
\hline
$\beta_1$ &  2.00 (1.97,2.03) & 2.00 (1.97,2.02)  & 2.00 (1.97,2.03)  & 2.00  (1.97,2.03)   \\
\hline
$\beta_2$  & 0.99  (0.77,1.17) & 0.98 (0.79,1.18)  & 0.99 (0.79,1.18)  &  0.97 (0.77,1.20)   \\
\hline
$\sigma_{\varepsilon}$  &  0.50 (0.46,0.53) & 0.50 (0.46,0.53)  & 0.50 (0.46,0.54)  & 0.50  (0.46,0.53)   \\
\hline
$\gamma_{\varepsilon}$  & 0.02  (-0.15,0.21) & 0.53 (0.35,0.78)  & 0.01 (-0.16,0.20)  & 0.52  (0.35,0.77)   \\
\hline
$\mu$   & -1.51  (-1.88,-1.06) &  -1.47 (-1.92,-1.13)  & -1.46 (-1.84,-1.12)  &  -1.44 (-1.79,-1.14)   \\
\hline
$\sigma$  &  0.48 (0.41,0.59) & 0.49 (0.41,0.57)  &  0.49 (0.43,0.56)  & 0.49  (0.43,0.57)   \\
\hline
$\gamma$  & -0.04  (-0.43,0.45) & -0.02 (-0.57,0.33)  &  0.51 (0.10,0.82)  & 0.52  (0.09,0.763)   \\
\hline
Median BF  & & & & \\ $\gamma_{\varepsilon}=0$ &  7.13   &  2$\times 10^{-21}$  &  7.04  &  1$\times10^{-19}$   \\
\hline
Median BF  & & & & \\ $\gamma=0$ &  2.45   &  2.81  & 0.40   &  0.29  \\
\hline
\end{tabular}
\caption{\small Simulation study III: Monte Carlo median, 95\% credible intervals of the median estimators using the two-piece normal model. BF are Bayes factors in favour of symmetry.}\label{table:SimulationTP}
\end{center}
\end{table}

\begin{table}[h!]
\begin{center}
\begin{tabular}[h]{|c|c|c|c|c|}
\hline
 & Scenario 1  & Scenario 2 & Scenario 3 & Scenario 4  \\
Parameter &  $\varepsilon_{ij}\sim \mbox{N}$ & $\varepsilon_{ij}\sim \mbox{TPN}$ & $\varepsilon_{ij}\sim \mbox{N}$  & $\varepsilon_{ij}\sim \mbox{TPN}$  \\
 & $u_i\sim \mbox{N}$  & $u_i\sim \mbox{N}$  & $u_i\sim \mbox{TPN}$  & $u_i\sim \mbox{TPN}$  \\
\hline
$\beta_1$ & 2.00 (1.97,2.03) &  2.00 (1.97,2.02)  &  2.00 (1.97,2.03)  & 2.00 (1.97,2.02)     \\
\hline
$\beta_2$ & 1.01 (0.78,1.18) & 0.99 (0.81,1.20)  & 0.99  (0.79,1.23)  & 0.98 (0.78,1.25)     \\
\hline
$\sigma_{\varepsilon}$ & 0.50 (0.46,0.54) & 0.53 (0.49,0.57)  & 0.50  (0.46,0.54)  & 0.53  (0.49,0.57)     \\
\hline
$\mu$  &  -1.51 (-1.65,-1.33) & -1.89 (-2.04,-1.73)  &  -1.89 (-2.08,-1.76)  &  -2.28 (-2.45,-2.14)     \\
\hline
$\sigma$ & 0.49 (0.41,0.58) & 0.50 (0.42,0.58)  &  0.52 (0.44,0.60)  &  0.52 (0.44,0.60)     \\
\hline
\end{tabular}
\caption{\small  Simulation study III: Monte Carlo median, quantile intervals of the median estimators using the Normal model.}\label{table:SimulationNTP}
\end{center}
\end{table}

\end{document}